\pdfoutput=1

\documentclass[11pt]{article}

\usepackage{acl}

\usepackage{times}
\usepackage{latexsym}

\usepackage[T1]{fontenc}

\usepackage[utf8]{inputenc}

\usepackage{microtype}

\usepackage{inconsolata}

\usepackage{graphicx}

\usepackage{booktabs}
\usepackage{amsmath} 
\usepackage{makecell}
\usepackage{colortbl}  
\usepackage{xcolor}
\usepackage{hhline}
\usepackage{enumerate}
\usepackage{tabularx}
\usepackage{tcolorbox}
\usepackage{multirow}
\usepackage{array}
\usepackage{courier}

\title{KeyInst: Keyword Instruction for Improving SQL Formulation in Text-to-SQL}

\author{Xiping Liu ,  Zhao Tan \\
        Jiangxi University of Finance and Economics}

\begin{document}
\maketitle
\begin{abstract}
Text-to-SQL parsing involves the translation of natural language queries (NLQs) into their corresponding SQL commands. A principal challenge within this domain is the formulation of SQL queries that are not only syntactically correct but also semantically aligned with the natural language input. However, the intrinsic disparity between the NLQ and the SQL poses a significant challenge. In this research, we  introduce \emph{Keyword Instruction (KeyInst)}, a novel method designed to enhance SQL formulation by Large Language Models (LLMs). KeyInst essentially provides guidance on pivotal SQL keywords likely to be part of the final query, thus facilitates a smoother SQL query formulation process. We explore two strategies for integrating KeyInst into Text-to-SQL parsing: a pipeline strategy and a single-pass strategy. The former first generates KeyInst for question, which are then used to prompt LLMs. The latter employs a fine-tuned model to concurrently generate KeyInst and SQL in one step. We developed \emph{StrucQL}, a benchmark specifically designed for the evaluation of SQL formulation. Extensive experiments on StrucQL and other benchmarks demonstrate that KeyInst significantly improves upon the existing Text-to-SQL prompting techniques.
\end{abstract}

\section{Introduction}
The task of Text-to-SQL parsing, which aims at translating natural language questions into executable SQL queries, has gained increasing attention in recent years, as it can help non-expert users quickly access information in the database without the need for technical background \cite{deng2021structure, yugrappa, rajkumar2022evaluating, ni2023lever}. 
Text-to-SQL parsing faces two main challenges: schema linking and SQL formulation. \emph{Schema linking} involves identifying the pertinent tables and columns in a database schema in response to an NLQ. \emph{SQL formulation} refers to generating SQL queries that are not only syntactically correct but also semantically aligned with the natural language input.

This paper primarily focuses on the challenge of SQL formulation. Currently, most Text-to-SQL prompting methods induce Large Language Models (LLMs) to generate the target SQL directly using In-context Learning (ICL) \cite{nan2023enhancing, pourreza2024din, tan2024enhancing}. However, the vast difference between natural language queries (NLQ) and SQL hinders precise query formulation. In previous works, the skeleton-aware decoder \cite{li2023resdsql} was proposed to alleviate this challenge by initially generating an SQL skeleton followed by the full query. An SQL skeleton is a basic framework of an SQL query consisting of SQL operators, without specific details such as column names, table names, or conditions.
Incorporating SQL skeleton in prompting has also proven to be effective \cite{gao2023text, guo2023prompting}. 
In this work, we also use the SQL structure as a central element in SQL formulation, with a particular emphasis on identifying key SQL operators. For instance, in translating the NLQ "List the customers' first and last names from the 10 least expensive invoices", accurately identifying \texttt{ORDER BY} and \texttt{LIMIT} is crucial for formulating the correct SQL query.

\begin{figure*}[!t]
    \centering
    \includegraphics[width=0.98\linewidth,scale=1]{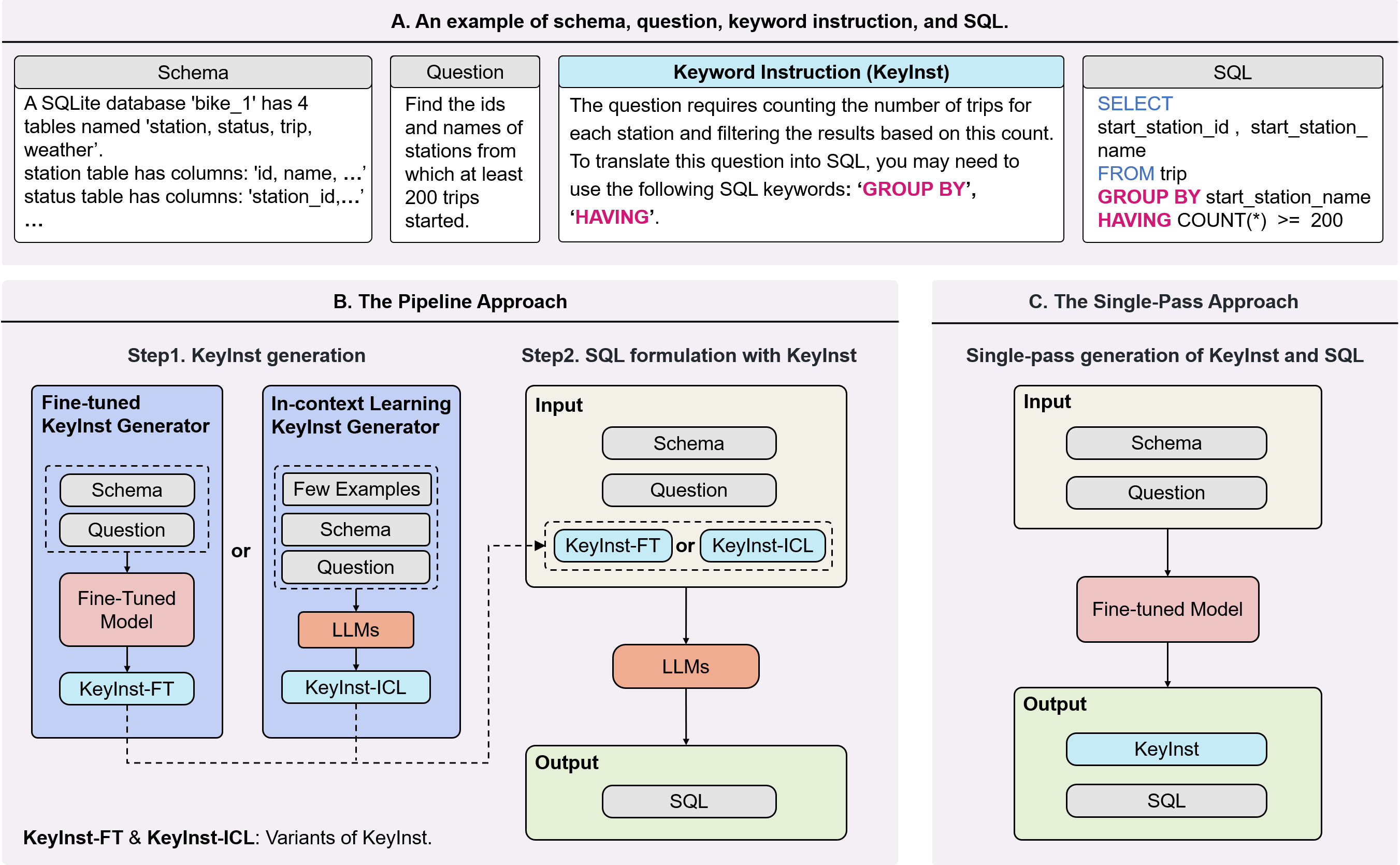}
    \caption{Graphical illustration of KeyInst and its applications: A. An example of schema, question, KeyInst, and SQL, B. The pipeline approach of KeyInst application, C. The single-pass approach of KeyInst application.}
    \label{fig: Graphical illustration of our methods}
\end{figure*}

We introduce \emph{Keyword Instruction (KeyInst)}, a novel method designed to enhance SQL formulation by LLMs. KeyInst essentially provides guidance on pivotal SQL keywords likely to be part of the final query. Recognizing that SQL queries corresponding to different NLQs require distinct keywords, KeyInst adapts dynamically to each query. An example of KeyInst in action is depicted in Figure~\ref{fig: Graphical illustration of our methods}A, demonstrating how it analyzes a given NLQ and deduces the critical SQL keywords. This strategy effectively narrows the gap between NLQ and SQL, facilitating a smoother SQL query formulation process.

While KeyInst significantly aids in SQL query formulation, further exploration is needed on its generation and integration into Text-to-SQL parsing. We present two approaches for KeyInst generation: a model fine-tuning method and an ICL-based method. The former fine-tunes a model to produce KeyInsts for specific queries, while the latter prompts LLMs to generate KeyInsts through ICL \cite{brown2020language}. For the application of KeyInst in Text-to-SQL tasks, we also investigate two strategies. The first strategy prompts LLMs to produce SQL queries with KeyInst. The second strategy is a fine-tuning strategy that generates SQL queries directly following KeyInst generation, treating KeyInst creation as a preliminary reasoning step.

To amalgamate KeyInst generation and application within Text-to-SQL, we introduce a two-fold strategy. The pipeline approach initially generates KeyInst using either the fine-tuned or ICL-based method, followed by prompting LLMs with the generated KeyInst, as illustrated in Figure~\ref{fig: Graphical illustration of our methods}B. Conversely, the single-pass approach employs a fine-tuned model to concurrently generate KeyInst and SQL in one step, as depicted in Figure~\ref{fig: Graphical illustration of our methods}C.


Several benchmarks, such as Spider \cite{yu2018spider} and Bird \cite{li2024can}, have been developed to assess Text-to-SQL systems. However, these benchmarks focus on overall parsing performance and lack mechanisms for isolating evaluations of semantic linking and SQL formulation. To specifically assess SQL formulation capabilities, a new benchmark called \emph{StrucQL (Structural Benchmark for Text-to-SQL)} has been developed, derived from Spider. In StrucQL, questions and schemas are simplified: questions explicitly mention schema items, and irrelevant tables and columns are omitted from the schema. This simplification makes schema linking straightforward, shifting the primary challenge to SQL formulation. Consequently, StrucQL serves as an effective tool for evaluating SQL formulation proficiency in Text-to-SQL systems.

KeyInst was assessed on StrucQL and other benchmarks, with outcomes indicating that keyword instructions are a valuable intermediary for Text-to-SQL parsing, whether applied independently or in conjunction with other techniques.


The main contributions of this work are summarized as follows:
\begin{itemize}
    \item We propose KeyInst, a keyword instruction tailored for each Text-to-SQL task, to alleviate SQL formulation challenge. We offer two approaches for integrating KeyInst into Text-to-SQL parsing: a pipeline strategy and a single-pass strategy.
    
    \item The StrucQL benchmark was developed to specifically assess the SQL formulation abilities of Text-to-SQL systems. By simplifying questions and schemas, StrucQL eliminates schema linking challenges, focusing evaluation on SQL formulation performance.
        
    \item Comprehensive experiments across various benchmarks were conducted. The findings demonstrate that KeyInst significantly improves upon the existing state-of-the-art Text-to-SQL prompting techniques, showcasing its effectiveness and potential.
\end{itemize}

\section{Methods}
\label{sec: Methods}
This paper introduces KeyInst to address the challenge of SQL formulation in Text-to-SQL parsing. The main idea is to analyze the NLQ to understand its intent, providing explicit guidance for SQL formulation by identifying essential keywords crucial for translating the NLQ into the target SQL. KeyInst is generated in real-time for each Text-to-SQL task.

We prepared over 6,200 KeyInst examples from the Spider training set, organized into a KeyInst set $S_{KeyInst} = \{(D_i, Q_i, K_i, S_i)\}$, where $D_i$ is the database schema, $Q_i$ is the question, $S_i$ is the SQL, and $K_i$ is the KeyInst.

Each KeyInst consists of two parts: question analysis and keyword suggestion, as shown in Figure \ref{fig: An example of the KeyInst.}. The question analysis is generated by prompting LLMs (see Appendix \ref{sec: Prompt towards Question Analysis in the KeyInst Set}). For keyword suggestions, we parse the SQL structure to identify all the keywords it uses, then filter out non-essential ones. Keywords are prioritized as follows: highest priority (\texttt{GROUP BY}, \texttt{HAVING}, \texttt{ORDER BY}, \texttt{LIMIT}, \texttt{EXCEPT}, \texttt{INTERSECT}, \texttt{UNION}, \texttt{WHERE}), second priority (\texttt{SELECT}, \texttt{FROM}). Lower priority keywords are only added if an SQL lacks higher priority keywords. Other Keywords(\texttt{JOIN}, \texttt{COUNT}, \texttt{IN}, and \texttt{others}) are excluded from the keyword suggestions. 
Without keyword prioritization, KeyInst would degrade into an SQL skeleton, which includes all the keywords of an SQL statement. More details about the skeleton can be found in Appendix \ref{sec: Comparsion of KeyInst and SQL skeleton.}.

\begin{figure}[!t]
    \centering
    \includegraphics[width=0.9\linewidth,scale=1]{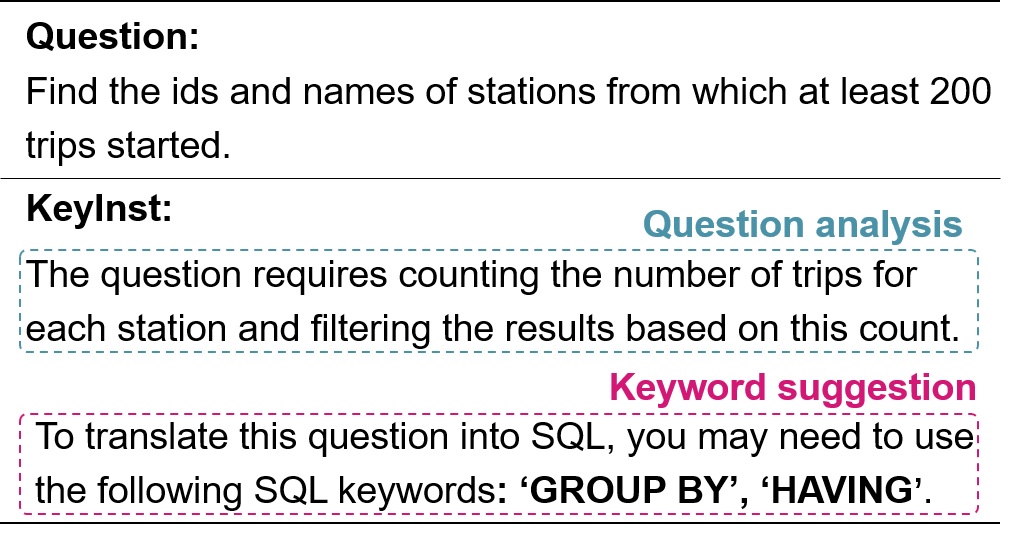}
    \caption{An example of the KeyInst.}
    \label{fig: An example of the KeyInst.}
\end{figure}

We implement the applications of KeyInst based on the KeyInst set $S_{KeyInst}$, detailed in the following sections.

\subsection{Pipeline Approach of KeyInst}
We propose the pipeline approach as one application of KeyInst, generating SQL in two steps. First, generating the tailored KeyInst for the each Text-to-SQL task, then prompt LLMs with the generated KeyInst to generate SQL query. This section details this application.

\subsubsection{Fine-tuned KeyInst Generator}
\label{sec: Fine-tuned KeyInst Generator}
One approach for KeyInst generation is to fine-tune a model to become a KeyInst generator. Using supervised fine-tuning, the input is a database shcema $D_i$ a question $Q_i$, and the target output is the corresponding KeyInst $K_i$. The primary objective is to minimize the following loss function:
\begin{equation}
 \min_{\theta} \frac{1}{N} \sum_{i=1}^N \mathcal{L} \left(\mathcal{M}_{\theta}\left(D_i, Q_i\right), K_i    \right),
\end{equation}
where $\mathcal{L}$ represents the loss related to the model’s next token prediction, comparing the predicted KeyInst with the actual ground truth.
This fine-tuned model, referred to as the \emph{fine-tuned KeyInst generator}, analyzes the question and generates a tailored KeyInst, called \emph{KeyInst-FT}, to prompt LLMs in SQL formulation.

\subsubsection{In-context Learning KeyInst Generator}
\label{sec: In-context KeyInst Generator}
Another apporach for KeyInst generation is to prompt LLMs with ICL to generate KeyInst. We select a few examples (i.e., demonstrations) from the KeyInst set $S_{KeyInst}$ to form a few-shot prompt for generating a tailored KeyInst for a Text-to-SQL task. Each example contains a database schema $D_i$, a question $Q_i$ and its corresponding KeyInst $K_i$. For each Text-to-SQL task, we select the top-$m$ most similar examples based on masked question similarity \cite{gao2023text} and combine them with the current database schema $D$ and question $Q$ to create a few-shot prompt. This few-shot prompt guides the LLMs to generate the tailored KeyInst, named \emph{KeyInst-ICL}, for the current question. The process can be formulated as:
\begin{equation}
\resizebox{\linewidth}{!}{ $
P_{\text {LLMs}}(y \mid x)=P\left(y \mid \operatorname{prompt}\left( \left(D, Q \right),\left\{\left(D_i, Q_i, K_i\right)\right\}_{i<=m}\right)\right),
$}
\end{equation}
where $x$ is the LLMs' input, including the current schema and question and the $m$ examples. The output $y$ is the expected KeyInst-ICL for the current question. This system is referred to as the \emph{in-context learning KeyInst generator}.

\subsubsection{SQL formulation with KeyInst}
SQL formulation follows KeyInst generation. A basic usage is to combine it with the database schema $D$, question $Q$, and KeyInst $K$ to construct a zero-shot prompt that guides LLMs to generate SQL. This can be formulated as:
\begin{equation}
\label{equ: SQL formulation with LLMs}
P_{\text {LLMs}}(y \mid x)=P\left(y \mid \operatorname{prompt}\left(D, Q, K\right)\right),
\end{equation}
where $x$ is the LLMs' input (i.e., the zero-shot prompt), and $y$ is the expected SQL output.

Notably, KeyInst functions as an instruction to enhance the SQL formulation capabilities of LLMs. It is highly extensible and can be effortlessly integrated with existing Text-to-SQL prompting methods. By appending KeyInst to these methods' prompts, their performance can be significantly improved. This will be analyzed in detail in \S \ref{sec: Results on General Benchmark}.

\subsection{Single-Pass Approach of KeyInst}
\label{sec: Fine-tuning with KeyInst}
We propose a single-pass approach for KeyInst as an alternative application method of KeyInst. In this approach, a fine-tuned model simultaneously generates both KeyInst and SQL in a single pass.  The generation of KeyInst serves as an initial reasoning step, and the fine-tuning process helps the model internalize this reasoning, thereby improving its ability to formulate SQL queries.

This involves supervised fine-tuning, where inputs are the database schema $D_i$ and the question $Q_i$, and targets are the KeyInst $K_i$ and the SQL statement $S_i$. The objective is to minimize the empirical loss:
\begin{equation}
 \min_{\theta} \frac{1}{N} \sum_{i=1}^N \mathcal{L} \left(\mathcal{M}_{\theta}\left(D_i , Q_i\right), \left(K_i, S_i \right)    \right),
\end{equation}
where $\mathcal{L}$ represents the loss related to the model’s next token prediction, comparing the predicted KeyInst and SQL with the actual ground truth.

The key difference between our fine-tuned model and a common Text-to-SQL fine-tuning model lies in the output. Our model first generates a KeyInst before generating the SQL, effectively reasoning about SQL formulation. Fine-tuning enables the model to remember this reasoning pattern, so it can spontaneously perform the reasoning when encountering a Text-to-SQL task.

\section{StrucQL: A Structural Benchmark for Text-to-SQL}
StrucQL is developed to allow researchers to swiftly and independently assess the SQL formulation capabilities of Text-to-SQL systems. Text-to-SQL errors can be categorized into semantic errors, which reflect schema linking capabilities, and structural errors, which pertain to SQL formulation skills.

However, widely-used Text-to-SQL benchmarks such as Spider \cite{yu2018spider} and Bird \cite{li2024can} focus on overall parsing performance and lack mechanisms for isolating evaluations of SQL formulation. Previous studies have relied on expensive manual evaluations to gauge structural performance \cite{ning2024insights}, underscoring the necessity for a specialized structural benchmark for Text-to-SQL systems.

SQL structural errors fall into two categories: (1) syntax errors, such as mismatched parentheses, which make SQL unexecutable, and (2) structural misalignments with the NLQ, such as inappropriate keyword usage. While LLMs can easily generate syntactically correct SQL due to extensive pre-training, the real challenge is ensuring structural alignment with the NLQ. StrucQL, therefore, focuses on evaluating this alignment.

We developed StrucQL by modifying the Spider dataset and utilizing GPT4\footnote{https://openai.com \label{gpt4}} for assistance. StrucQL comprises 1050 examples and covers 7 types of SQL operation. Each example is dedicated to a single operation type. To mitigate schema linking difficulties in the Text-to-SQL task, we implemented schema simplification. This process reduces schema linking errors, thereby providing a clearer assessment of SQL formulation. Specifically, we replaced schema-related terms in the original NLQs with the corresponding table and column names, and filtered out tables and columns from the database schema that were irrelevant to the questions (see Appendix \ref{sec: An Example of Schema Simplification} for an example).

\begin{table}[t]
    \centering
    \resizebox{\linewidth}{!}{
    \begin{tabular}{lccccc}
    \toprule[1pt]
    \textbf{Type} & \textbf{Gemma-7B} & \textbf{Llama3-8B} & \textbf{Llama3-70B} & \textbf{Claude3} & \textbf{GPT4}\\
    \midrule[0.5pt]
    \multicolumn{6}{c}{\textit{Original question and schema (Input)}}\\
    \midrule[0.5pt]

    GROUP BY & 47.3  & 63.3   & 75.3  & 72.0    & \textbf{75.3}    \\
    HAVING   & 50.0    & 68.0   & 80.0  & 83.3    & \textbf{86.0}    \\
    ORDER BY & 52.7    & 76.0    & 82.0    & 90.7   & \textbf{92.7}     \\
    LIMIT    & 44.0    & 54.0   & 66.0   & 74.7   & \textbf{80.0}   \\
    EXCEPT   & 36.7      & 59.3    & 63.3     & 68.7    & \textbf{71.3}      \\
    INTERSECT& 30.0      & 48.7    & 59.3     & 69.3   & \textbf{70.7}     \\
    UNION    & 17.3     & 37.3   & 46.7   & 53.3     & \textbf{56.7}   \\
    \midrule[0.5pt]
    Overall  & 39.7    & 58.1    & 67.5    & 73.1    & \textbf{76.1}  \\
    
    \midrule[0.5pt]
    \midrule[0.5pt]
    
    \multicolumn{6}{c}{\textit{Schema-simplified question and schema (Input)}}\\
    \midrule[0.5pt]
    GROUP BY & 58.7    & 69.3    & \textbf{78. 0}  & 73.3   & 76.7   \\
    HAVING   & 52.7    & 74.0    & 82.0            & 85.3   & \textbf{86.0}   \\
    ORDER BY & 66.7    & 74.7    & 90.7            & 92.7    & \textbf{94.0}    \\
    LIMIT    & 51.3    & 76.0    & 80.7    & 77.3   & \textbf{88.7}   \\
    EXCEPT   & 41.3    & 63.3    & 64.7      & 70.7    &\textbf{ 72.7}    \\
    INTERSECT& 42.7    & 57.3    & 61.3   & \textbf{73.3}   & 72.7   \\
    UNION    & 30.7    & 38.7    & 47.3  & 55.3   & \textbf{57.3}  \\
    \midrule[0.5pt]
    Overall  & 49.1   & 63.9   & 72.0    & 75.4   & \textbf{78.3} \\
    
    \bottomrule[1pt]
    \end{tabular}
    }
    \caption{The results of execution accuracy for all compared models on StrucQL.}
    \label{tab:StrucQL result}
\end{table}

Table \ref{tab:StrucQL result} presents the performance of various LLMs on StrucQL in a zero-shot scenario. To assess the impact of schema simplification on the Text-to-SQL task, we compared the results of models using original inputs with those using schema-simplified inputs. Overall, schema simplification improved execution accuracy. Larger models exhibited smaller improvements: GPT-4's accuracy increased by 2.2\%, whereas Gemma-7B's accuracy rose by 9.4\%.
Additionally, we observed a significant performance gap for set operation types (\texttt{EXCEPT}, \texttt{INTERSECT}, \texttt{UNION}) compared to other types. This disparity may be due to the relative infrequency of set operations, leading to less representation in the models' training datasets. Moreover, schema simplification did not significantly enhance execution accuracy for these types, suggesting that their primary challenges are not related to schema linking issues.

To the best of our knowledge, StrucQL is the first effective tool designed to evaluate SQL formulation proficiency in Text-to-SQL systems. It offers deeper insights into the methodologies of these systems. By focusing on various types of SQL operations, StrucQL allows for a targeted evaluation of specific operations, helping to identify particular strengths and weaknesses in SQL formulation. Ultimately, this leads to more robust and accurate Text-to-SQL systems, enhancing database interactions.

\section{Experiments}
In this section, we systematically assess the effectiveness of KeyInst. Our evaluation centers on two primary aspects: (1) comparing the performance of different applications of KeyInst, and (2) examining the performance improvements when KeyInst is integrated with current state-of-the-art (SOTA) Text-to-SQL prompting methods.

\subsection{Setup}
\textbf{Models} \hspace{0em} We selected five LLMs for our experiments: Gemma-7B-It \cite{team2024gemma} (Gemma-7B) , Llama-3-8B-Instruct\footnote{https://github.com/meta-llama/llama3 \label{llama3}} (Llama3-8B) , Llama-3-70B-Instruct\footref{llama3} (Llama3-70B), Claude-3-Opus-20240229\footnote{https://claude.ai} (Claude3), and GPT-4-Turbo-2024-04-09 \footref{gpt4} (GPT4). 

~\\\textbf{Hyperparameters} \hspace{0em} For fine-tuning method, the Llama3-8B model is trained on Nvidia Tesla A100 GPUs, employing a batch size of 32 with a learning rate of 1*e-5. For the prompting method, we perform greedy decoding at a temperature of $\tau$ = 0 to ensure reproducible results.

~\\\textbf{Benchmarks} \hspace{0em} We used the following benchmarks: StrucQL, Spider \cite{yu2018spider}, and Bird \cite{li2024can}.
StrucQL, introduced in this paper, evaluates the SQL formulation of Text-to-SQL systems.
Spider is a large-scale benchmark with 8,000 training samples and 1,034 development samples across multiple databases.
The BIRD dataset features 12,751 question-SQL pairs, covering 95 large databases across 37 professional fields.

~\\\textbf{Metrics} \hspace{0em} We use execution accuracy (EX) to evaluate different methods. This metric compares the execution output of the predicted SQL query with that of the ground truth SQL query on same database instances.

~\\\textbf{Baselines} \hspace{0em} 
We selected three SOTA Text-to-SQL prompting methods as the baselines.

\noindent (1) \textit{DIN-SQL}  \cite{pourreza2024din}: This pipeline prompting method that involves schema linking, difficulty classification, SQL generation, and SQL self-correction.

\noindent (2) \textit{DAIL-SQL}  \cite{gao2023text}: An efficient few-shot prompting method that selects demonstrations based on similarity of masked question and SQL skeleton. We use DAIL-SQL with 8 shots.

\noindent (3) \textit{SC-SQL} \cite{tan2024enhancing}: This method integrates multiple Text-to-SQL reasoning paths and selects the best candidate result from these paths.

~\\\textbf{Our methods} \hspace{0em} We introduce KeyInst and apply it using two approaches: the pipeline approach and the single-pass approach. The specific implementations are:

\noindent (1) \textit{KeyInst-FT}: A variant of KeyInst generated by the fine-tuned KeyInst generator (\S \ref{sec: Fine-tuned KeyInst Generator}), used to prompt LLMs to generate SQL. Specifically, we fine-tuned a Llama3-8B model as the fine-tuned KeyInst generator.

\noindent (2) \textit{KeyInst-ICL}: Another variant of KeyInst generated by the in-context learning KeyInst generator (\S \ref{sec: In-context KeyInst Generator}). For each Text-to-SQL task, 6 examples are retrieved based on masked question similarity \cite{gao2023text} to create a few-shot prompt, which is then used to guide GPT4 in generating KeyInst-ICL.

\noindent (3) \textit{KeyLla}: As described in \S \ref{sec: Fine-tuning with KeyInst}, we fine-tuned a Llama3-8B model with the KeyInst set, resulting in the KeyLla model. This model can perform KeyInst reasoning first and then generate SQL.

\subsection{Results on StrucQL}
We introduce StrucQL as a benchmark designed to evaluate the performance of Text-to-SQL systems in the challenge of SQL formulation. In this section, we analyze the performance of various applications of KeyInst on the StrucQL benchmark.

\subsubsection{Comparison of KeyInst Applications}
\label{sec: Comparison of Fine-tuning and Prompting}
we compared two applications of KeyInst: the pipeline approach and the single-pass approach.
The KeyLla, a Llama3-8B model fine-tuned with KeyInst, represents the single-pass approach.  To ensure a fair comparison, we also used the Llama3-8B model within the pipeline to generate SQL. We utilized two variants of KeyInst: KeyInst-FT and KeyInst-ICL. 
Despite their different generation methods, both variants serve as part of the prompt to guide the Llama3-8B in SQL formulation. The results are detailed in Table \ref{tab: Comparison of Fine-tuning and Prompting}.

\begin{table}[t]
    \centering
    \resizebox{0.8\linewidth}{!}{
    \begin{tabular}{lccc}
    \toprule[1pt]
    \multirow{2}{*}{\textbf{Type}}  &  \textbf{Single-Pass}     & \multicolumn{2}{c}{\textbf{Pipeline}} \\
    \cmidrule(lr){2-2} \cmidrule(lr){3-4}
                             &  \textbf{KeyLla} & \textbf{KeyInst-FT} &  \textbf{KeyInst-ICL}\\
    \midrule[0.5pt]
    GROUP BY & \textbf{80.7}    &  74.5    &    79.3  \\
    HAVING   & \textbf{85.3}   &  80.0    &    82.0  \\
    ORDER BY & \textbf{90.7}  &  88.0    &    87.3  \\
    LIMIT    & 83.3    &  82.7    &    \textbf{86.7}  \\
    EXCEPT   & \textbf{78.0}    &  69.3    &    73.3  \\
    INTERSECT& \textbf{82.0}    &  81.3    &    80.7  \\
    UNION    & 53.3    &  \textbf{62.0}    &    54.0  \\
    \midrule[0.5pt]
    Overall  & \textbf{79.1}    &  76.9    &    76.8  \\
    \bottomrule[1pt]
    \end{tabular}
    }
    \caption{Execution accuracy (EX) of KeyInst applications on StrucQL. KeyLla is a fine-tuned Llama3-8B model. KeyInst-FT and KeyInst-ICL are variants of KeyInst, used to prompt LLMs (here, the Llama3-8B) for SQL formulation.}
    \label{tab: Comparison of Fine-tuning and Prompting}
\end{table}

Table \ref{tab: Comparison of Fine-tuning and Prompting} demonstrates that the single-pass approach (i.e., KeyLla) outperforms the pipeline approach (i.e., KeyInst-FT and KeyInst-ICL) in terms of performance. KeyLla effectively internalizes KeyInst's reasoning for SQL formulation, leading to superior results.
While the single-pass approach achieved the best performance, the pipeline approach can also achieve comparable results. 
Additionally, the pipeline approach have the advantage of being able to leverage more powerful LLMs, such as GPT4. Achieving similar performance through single-pass approach would be significantly more costly.

In conclusion, both single-pass and pipeline are effective for addressing the SQL formulation challenge in Text-to-SQL tasks.
If budget allows, single-pass approach of KeyInst can achieve better performance compared to pipeline. However, the advantage of pipeline lies in their cost-effectiveness and flexibility. Pipeline can be combined with more powerful LLMs without requiring extensive computational resources and time-consuming training processes.

\subsubsection{Comparison of KeyInst-FT and KeyInst-ICL}

\begin{table}[t]
    \centering
    \resizebox{0.6\linewidth}{!}{
    \begin{tabular}{lcc}
    \toprule[1pt]
    \textbf{Models} & \textbf{Prompting} & \textbf{EX}  \\
    \midrule[0.5pt]
    \multirow{2}{*}{Llama3-8B}  & KeyInst-ICL & 76.8   \\
                               & KeyInst-FT & \textbf{76.9}  \\ 
    \midrule[0.5pt]
    \multirow{2}{*}{Llama3-70B}  & KeyInst-ICL & 80.5  \\
                                & KeyInst-FT & \textbf{83.2}  \\ 
    \midrule[0.5pt]            
    \multirow{2}{*}{GPT4} & KeyInst-ICL & 82.3   \\
    & KeyInst-FT & \textbf{84.3}   \\ 
    \bottomrule[1pt]
    \end{tabular}
    }
    \caption{Execution accuracy (EX) of various LLMs using KeyInst-FT and KeyInst-ICL on StrucQL.}
    \label{tab: Comparison of KeyInst-FT and KeyInst-IC}
\end{table}

Table \ref{tab: Comparison of KeyInst-FT and KeyInst-IC} presents the performance of KeyInst-FT and KeyInst-ICL on more powerful LLMs. Using KeyInst-FT to prompt GPT4 achieves the highest EX result at 84.3\%. 
This outcome demonstrates the advantage of the pipeline approach, which can achieve excellent performance with low computational resources by leveraging the powerful natural language processing capabilities of LLMs.

Table \ref{tab: Comparison of KeyInst-FT and KeyInst-IC} also highlights the superior performance of KeyInst-FT over KeyInst-ICL. KeyInst-ICL provides overly detailed keyword suggestions, such as \texttt{AVG}, \texttt{COUNT}, \texttt{JOIN}, and \texttt{IN} (see Appendix \ref{sec: Examples of KeyInst-FT and KeyInst-IC} for examples). These keywords, defined as the lowest priority in \S \ref{sec: Methods}, are not expected to appear in KeyInst. Excessive detail can hinder LLMs' Text-to-SQL performance \cite{tai2023exploring, tan2024enhancing}, which may explain KeyInst-ICL's slightly poorer results. Additionally, KeyInst-FT is generated by a Llama3-8B model fine-tuned on over 6200 KeyInst data points, while KeyInst-ICL is generated by prompting GPT-4 with a 6-shot prompt. Although GPT4 is more powerful, fine-tuning enables the fine-tuned KeyInst generator to better capture the relationship between NLQ and KeyInst, thereby generating KeyInsts more suitable for the Text-to-SQL task.

\subsubsection{Ablation Study}
Table \ref{tab: ablation study} presents the results of the ablation study. Each KeyInst consists of two parts: question analysis and keyword suggestion (see Figure \ref{fig: An example of the KeyInst.}). The question analysis explains the NLQ of the current Text-to-SQL task, while the keyword suggestion provides potential SQL keywords for the current Text-to-SQL task. To assess the contribution of each component, we compared single-pass and pipeline approaches both with and without these parts. For the single-pass approach, we split the training data accordingly, and for the pipeline approach, we separated the KeyInst components when prompting the LLMs. The results in Table \ref{tab: ablation study} indicate that the keyword suggestion plays a more significant role in the effectiveness of KeyInst.

\begin{table}[t]
    \centering
    \resizebox{0.6\linewidth}{!}{
    \begin{tabular}{lc}
    \toprule[1pt]
    \textbf{Methods} & \textbf{EX}  \\
    \midrule[0.5pt]
    \multicolumn{2}{c}{\textit{The single-pass approach}}\\
    \midrule[0.5pt]
    KeyLla    &   79.1        \\
    \midrule[0.5pt]
    w/o question analysis    &     72.9  \\
    w/o keyword suggestion   &    68.4  \\
    \midrule[0.5pt]
    \midrule[0.5pt]
    \multicolumn{2}{c}{\textit{The pipeline approach}}\\
    \midrule[0.5pt]   
    KeyInst-FT + GPT4 & 84.3  \\
    \midrule[0.5pt]
    w/o question analysis & 83.4  \\
    w/o keyword suggestion & 81.5   \\
    \bottomrule[1pt]
    \end{tabular}
    }
    \caption{Ablation study.}
    \label{tab: ablation study}
\end{table}

\subsubsection{Results of Baselines}
Table \ref{tab: Results of baselines} presents the performance of baseline methods (DIN-SQL \cite{pourreza2024din}, DAIL-SQL \cite{gao2023text}, SC-SQL \cite{tan2024enhancing}) and our KeyInst-FT method on StrucQL. The results indicate that while these baseline methods, as SOTA Text-to-SQL prompting approaches, achieve commendable results on well-known benchmarks (e.g., Spider and Bird), there is still room for improvement in SQL formulation capabilities, particularly in set operations (\texttt{EXCEPT}, \texttt{INTERSECT}, \texttt{UNION}), where our method excels.
We believe that explicitly mentioning SQL keywords relevant to the current Text-to-SQL task in the prompt is crucial for enhancing the LLMs' SQL formulation performance. This is supported by DAIL-SQL's strong performance, which is attributed to its consideration of SQL skeleton similarity when constructing few-shot prompts, thus the prompt may contain important SQL keywords that are relevant to the current task.

\begin{table}[t]
    \centering
    \resizebox{\linewidth}{!}{
    \begin{tabular}{lcccc}
    \toprule[1pt]
    \textbf{Type} & \textbf{DIN-SQL} & \textbf{DAIL-SQL} & \textbf{SC-SQL} &  \textbf{KeyInst-FT}\\
    \midrule[0.5pt]
    GROUP BY & 75.3    & 77.3   & 76.7    & \textbf{78.7}       \\
    HAVING   & 85.3    & 85.3   & 84.0    & \textbf{86.0}       \\
    ORDER BY & \textbf{94.7}    & 94.0   & 93.3    & 94.0       \\
    LIMIT    & 88.7    & 89.3   & 88.0    & \textbf{89.3}       \\
    EXCEPT   & 75.3    & 82.0   & 78.7    & \textbf{83.3}        \\
    INTERSECT& 75.3    & 84.0   & 79.7    & \textbf{85.3}       \\
    UNION    & 62.0    & 62.7   & 65.3    & \textbf{73.3}       \\
    \midrule[0.5pt]
    Overall  & 79.5    & 82.1   & 81.0    & \textbf{84.3}        \\
    \bottomrule[1pt]
    \end{tabular}
    }
    \caption{Execution accuracy (EX) of GPT4 using baselines and KeyInst-FT on StrucQL.}
    \label{tab: Results of baselines}
\end{table}

\subsection{Results on General Benchmark}
\label{sec: Results on General Benchmark}

\begin{table}[t]
    \centering
    \resizebox{0.6\linewidth}{!}{
    \begin{tabular}{lcc}
    \toprule[1pt]
    \textbf{Methods} & \textbf{Spider} & \textbf{Bird}   \\
    \midrule[0.5pt]
    Zero-shot       & 77.9  &  43.6    \\
    DIN-SQL         & 85.1  &  50.7     \\
    DAIL-SQL        & 83.1  &  \textbf{54.8}    \\
    SC-SQL       & \textbf{86.2}  &  53.3    \\
    \midrule[0.5pt]
    KeyInst-FT      &  82.8    &  50.1      \\
    + DIN-SQL       &  86.8    &  54.5      \\
    + DAIL-SQL      &  85.2    &  \textbf{58.0}     \\
    + SC-SQL        &  \textbf{87.6}    &  56.6      \\
    \bottomrule[1pt]
    \end{tabular}
    }
    \caption{Execution accuracy (EX) of GPT4 on the Spider dev and Bird dev.}
    \label{tab: Results on General Benchmarks}
\end{table}

We also evaluated KeyInst on well-known benchmarks, such as Spider and Bird. KeyInst is instruction and can be easily integrated with existing Text-to-SQL prompting methods by appending KeyInst to their prompts (see Appendix \ref{sec: The usage of KeyInst} for examples). We used the GPT4 model to assess the performance of baseline methods with KeyInst, with results shown in Table \ref{tab: Results on General Benchmarks}.

We conducted experiments using KeyInst-FT (a variant of KeyInst). When used alone, KeyInst serves as a zero-shot prompting method. While it performs well compared to standard zero-shot methods, it does not match the current these SOTA Text-to-SQL prompting methods because it specifically addresses the SQL formulation challenge and does not focus on the schema linking challenge. However, this issue is easily resolved when combined with SOTA methods, which handle schema linking while KeyInst focuses on SQL formulation.

Table \ref{tab: Results on General Benchmarks} demonstrates significant performance improvements in baseline methods after incorporating KeyInst, highlighting KeyInst's effectiveness in SQL formulation. Additionally, KeyInst's ease of integration with prompting methods makes it a valuable tool for advancing prompt-based Text-to-SQL research.

\subsection{Discussion}
\textbf{How to choose the application method for KeyInst?} \hspace{0em} 
We propose two applications for KeyInst: single-pass and pipeline. When computational resources are abundant, the single-pass approach, which involves fine-tuning a model with KeyInst can maximize its SQL formulation capabilities. However, since computational resources are often limited, the pipeline approach becomes more advantageous as it can leverage more powerful models without extensive training.
Therefore, we believe that the pipeline approach for KeyInst deserves more attention. Within this approach, fine-tuning a KeyInst generator (if resources allow) can produce more effective KeyInsts than using the in-context learning KeyInst generator.

~\\\textbf{How to use KeyInst?} \hspace{0em} 
Due to KeyInst's lightweight design, KeyInst offers strong compatibility, especially evident in the prompting method. We do not recommend relying solely on KeyInst to solve Text-to-SQL tasks, as these tasks often also encounter the challenge of schema linking. We advocate for the integration of KeyInst with other Text-to-SQL prompting methods. This integration is straightforward because KeyInst is presented as an instruction within a prompt. The excellent compatibility of KeyInst holds significant potential for future research.

\section{Relate Work}

~\\\textbf{SQL formulation} \hspace{0em}
Previous works typically propose well-designed decoders to address SQL formulation challenge. \cite{wang2020rat, cai2021sadga, qi2022rasat}. RESDSQL \cite{li2023resdsql} introduces a skeleton-aware decoder that first generates an SQL skeleton and then fills the slots, proving to be very effective. A new trend involves prompting LLMs \cite{chen2023teaching, liu2023comprehensive}, focusing on task decomposition, or selecting demonstrations for few-shot prompts. DIN-SQL \cite{pourreza2024din} uses a pipeline to sequentially address schema linking and SQL formulation, while SC-SQL \cite{tan2024enhancing} emphasizes result consistency \cite{wang2022self}. DAIL-SQL \cite{gao2023text} selects demonstrations based on masked question and SQL skeleton similarity. Nan \cite{nan2023enhancing} and Guo \cite{guo2023prompting} propose similar methods. These approaches rely on implicit information in demonstrations, leading to suboptimal SQL formulation. In contrast, KeyInst explicitly guides LLMs to use specific SQL keywords through tailored instructions.

~\\\textbf{Benchmarks} \hspace{0em}
Popular benchmarks like Spider \cite{yu2018spider}, and Bird \cite{li2024can} evaluate comprehensive Text-to-SQL capabilities. More challenging datasets like Spider-Syn \cite{gan2021towards}, Spider-DK \cite{gan2021exploring}, and Ambiguity \cite{bhaskar2023benchmarking} focus on schema linking. However, the field lacks a benchmark for SQL formulation performance. Therefore, we propose StrucQL to help researchers assess the SQL formulation capabilities of Text-to-SQL systems.

\section{Conclusion}
This paper introduces KeyInst, a dynamic instruction method explicitly highlighting essential SQL keywords likely to be included in the target SQL query. We explore two approaches for integrating KeyInst into Text-to-SQL parsing: the pipeline approach and the single-pass approach.
In the pipeline approach, KeyInst is used to prompt LLMs. In contrast, the single-pass approach involves fine-tuning a model with KeyInst.
Our results indicate that, for models of the same size, the single-pass approach outperforms the pipeline approach. However, the pipeline approach excels in flexibility, easily integrating with more powerful LLMs to achieve superior performance.
Due to KeyInst's lightweight design, KeyInst integrates seamlessly with existing Text-to-SQL prompting methods, enhancing their performance. This compatibility suggests promising potential for future research in Text-to-SQL prompting.

\section*{Limitations}
In this paper, we made an effort to demonstrate the effectiveness of KeyInst, but there are still some limitations that need to be noted:
First and foremost, we acknowledge that KeyInst, designed for the SQL formulation challenge, offers limited assistance for the schema linking challenge. Whether this instruction-based method can be effectively used to address the schema linking challenge requires further exploration in the future.
Second, when discussing applications for KeyInst (\S \ref{sec: Comparison of Fine-tuning and Prompting}), due to budget constraints, we conducted experiments only on Llama3-8B. We are uncertain about the performance of the single-pass approach on larger models.
Third, prompting with KeyInst has shown excellent compatibility, as it can be combined with other prompting methods and enhance their performance. However, for fine-tuning with KeyInst, it remains unclear whether using KeyInst for fine-tuning existing Text-to-SQL models \cite{pourreza2024dtssql, li2024CodeS} will improve their performance. This requires further investigation in future research.

\bibliography{custom}

\appendix
\onecolumn

\section{Prompt of Question Analysis}
\label{sec: Prompt towards Question Analysis in the KeyInst Set}
In \S \ref{sec: Methods}, we constructed a KeyInst set, where the KeyInsts were pre-prepared. For the question analysis part of the KeyInsts, we used the following prompt to guide the GPT4 model. It is a few-shot prompt containing 7 demonstrations. The prompt is:\\

\noindent Please analyse the following natural language query.\\
Natural language query: Please show the different statuses of cities and the average population of cities with each status.\\
Analysis: The question is asking for a list of different statuses of cities and the average population for cities within each status. This requires grouping the cities by their status and calculating the average population for each group.\\

\noindent Please analyse the following natural language query.\\
Natural language query: What is the average longitude of stations that never had bike availability more than 10?\\
Analysis: The question is looking to calculate the average longitude of bike stations where the number of available bikes never exceeded 10. This requires filtering out stations based on a condition applied to their bike availability data. \\

\noindent Please analyse the following natural language query.\\
Natural language query: List the writers of the books in ascending alphabetical order. \\
Analysis: The question is asking to retrieve a list of writers from the book table and sort them in ascending alphabetical order. This requires selecting the Writer column and ordering the results.\\

\noindent Please analyse the following natural language query.\\
Natural language query: List the publisher of the publication with the highest price.\\
Analysis: The question is asking to identify the publisher of the publication that has the highest price. This requires sorting the publications by price in descending order and selecting the top result.\\

\noindent Please analyse the following natural language query.\\
Natural language query: Show ids for all employees who don't have a certificate.\\
Analysis: The question is asking for the IDs of employees who do not possess any certificates. This requires comparing two sets of data: one from the Employee table and one from the Certificate table, and then finding the difference between these two sets.\\

\noindent Please analyse the following natural language query.\\
Natural language query: Show names for all employees who have certificates on both Boeing 737-800 and Airbus A340-300.\\
Analysis: The question is looking for the names of employees who hold certificates for both the Boeing 737-800 and the Airbus A340-300 aircraft. This requires identifying employees who have certificates for both aircraft types and then retrieving their names.\\

\noindent Please analyse the following natural language query.\\
Natural language query: Find courses that ran in Fall 2009 or in Spring 2010.\\
Analysis: The question is looking for courses that were offered either in the Fall semester of 2009 or in the Spring semester of 2010. This requires filtering records based on specific conditions for both the semester and the year.\\

\section{Comparsion of KeyInst and SQL skeleton.}
\label{sec: Comparsion of KeyInst and SQL skeleton.}
In \S \ref{sec: Methods}, we assigned different priorities to SQL keywords and considered which keywords should be included in KeyInst's keyword suggestions based on these priorities. This approach was taken because general keywords (e.g., \texttt{JOIN}, \texttt{IN}, \texttt{COUNT}) do not directly reflect the query intent corresponding to NLQs. Instead, overly detailed information can increase the burden on the KeyInst generator and potentially affect the output of LLMs.

\begin{figure}[ht]
	\centering
	\includegraphics[width=0.9\linewidth,scale=1.00]{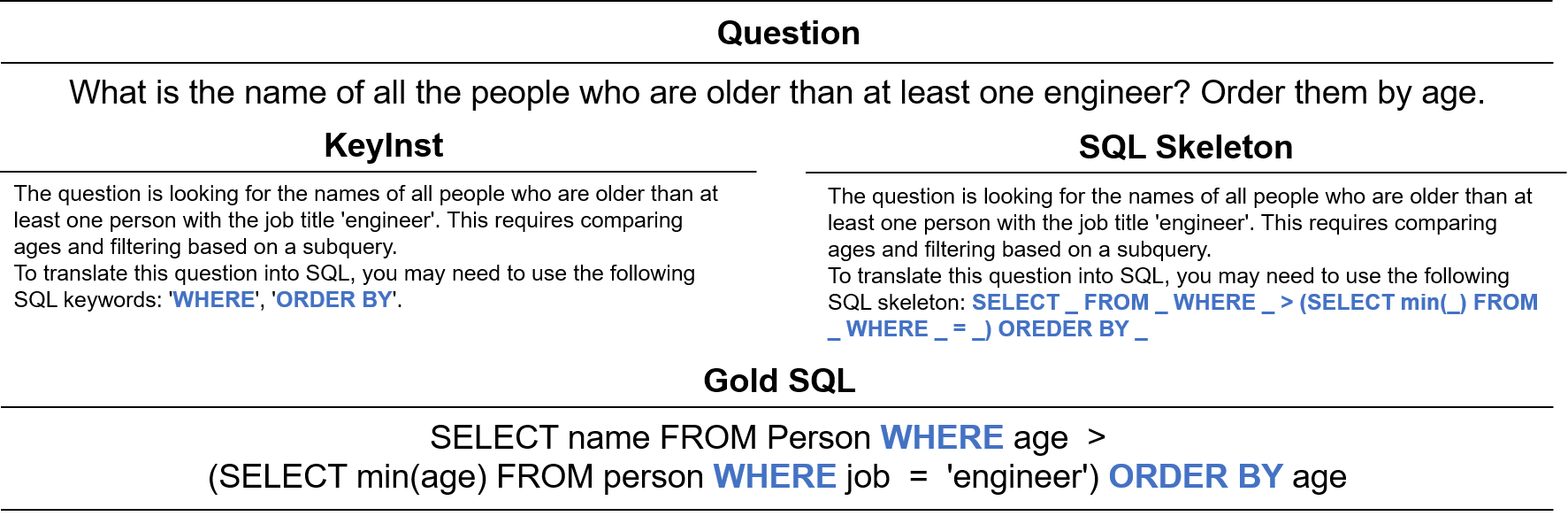}
	\caption{Examples of KeyInst-FT and SQL skeleton.}
    \label{fig: Comparsion of KeyInst and SQL skeleton}
\end{figure}

We conducted a comparative experiment where we did not set keyword priorities. In this scenario, KeyInst degraded into an SQL skeleton, as illustrated in Figure \ref{fig: Comparsion of KeyInst and SQL skeleton}. The details of the comparative experiment are as follows: we replaced the KeyInst in the KeyInst set with SQL skeletons and fine-tuned the Llama3-8B model to become an SQL skeleton generator. This generator produces an SQL skeleton for each Text-to-SQL task, and this generated SQL skeleton is then used as part of the prompt to guide the LLMs in generating SQL.

The results in Table \ref{tab: Comparsion of SQL skeleton and KeyInst.} show that using the skeleton is significantly less effective than using KeyInst-FT(a version of KeyInst). Although the skeleton (Figure \ref{fig: Comparsion of KeyInst and SQL skeleton}) appears more specific on the surface compared to KeyInst, directly deriving an SQL skeleton from an NLQ is not easy. This often leads to unexpected errors, which can mislead the LLMs when generating SQL.

\begin{table}[ht]
    \centering
    \resizebox{0.3\linewidth}{!}{
    \begin{tabular}{lcc}
    \toprule[1pt]
    \textbf{Type} & \textbf{Skeleton} & \textbf{KeyInst-FT}  \\
    \midrule[0.5pt]  
    GROUP BY &  77.3   & \textbf{78.7} \\
    HAVING   &  85.3   & \textbf{86.0}   \\
    ORDER BY &  92.7   & \textbf{94.0}   \\
    LIMIT    &  87.3   & \textbf{89.3}   \\
    EXCEPT   &  76.0   & \textbf{83.3}   \\
    INTERSECT&  78.7   & \textbf{85.3}   \\
    UNION    &  60.0  & \textbf{73.3 }    \\
    \midrule[0.5pt]
    Overall  & 79.6   & \textbf{84.3} \\
    \bottomrule[1pt]
    \end{tabular}
    }
    \caption{Execution accuracy (EX) of GPT4 using SQL skeleton and KeyInst-FT on StrucQL.}
    \label{tab: Comparsion of SQL skeleton and KeyInst.}
\end{table}

\section{An Example of Schema Simplification.}
\label{sec: An Example of Schema Simplification}
\begin{figure}[ht]
	\centering
	\includegraphics[width=0.8\linewidth,scale=1.00]{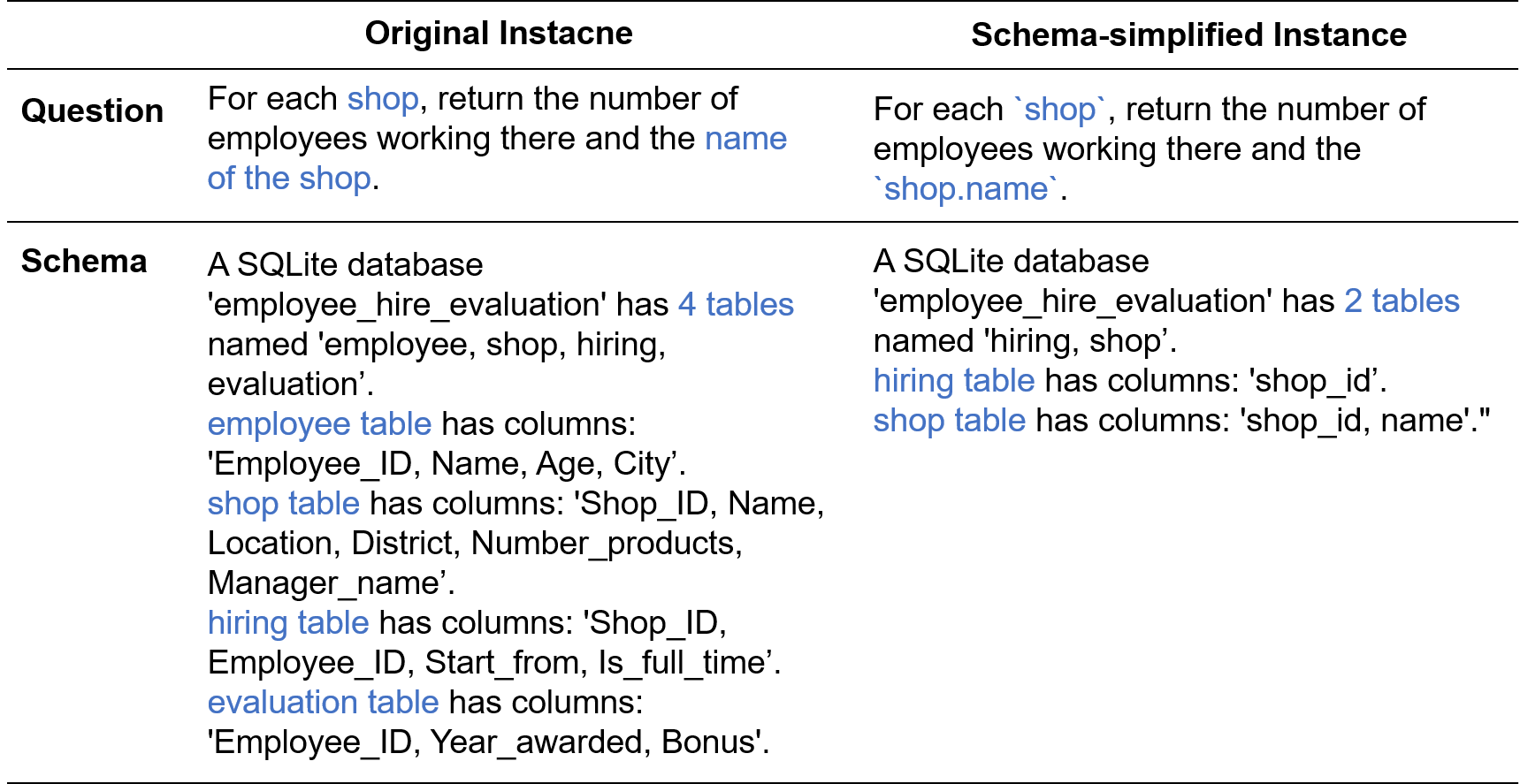}
	\caption{An example of Schema-Simplifed question and schema.}
    \label{fig: An Example of Schema Simplification}
\end{figure}

We constructed the StrucQL benchmark to intuitively evaluate a Text-to-SQL system's SQL formulation performance by minimizing schema linking's impact. This is achieved through schema simplification, which aims to reduce the complexity of schema linking in Text-to-SQL tasks, thereby decreasing the errors caused by incorrect schema links. Figure \ref{fig: An Example of Schema Simplification} provides an example of the schema-simplified question and schema. Specifically, we marked and modified schema-related words in the question. For instance, in Figure \ref{fig: An Example of Schema Simplification}, 'name of the shop' was changed to 'shop.name'. Additionally, we filtered out tables and columns from the database schema that are irrelevant to the current question.

\section{Examples of KeyInst-FT and KeyInst-ICL}
\label{sec: Examples of KeyInst-FT and KeyInst-IC}
\begin{figure}[!h]
	\centering
	\includegraphics[width=0.8\linewidth,scale=1.00]{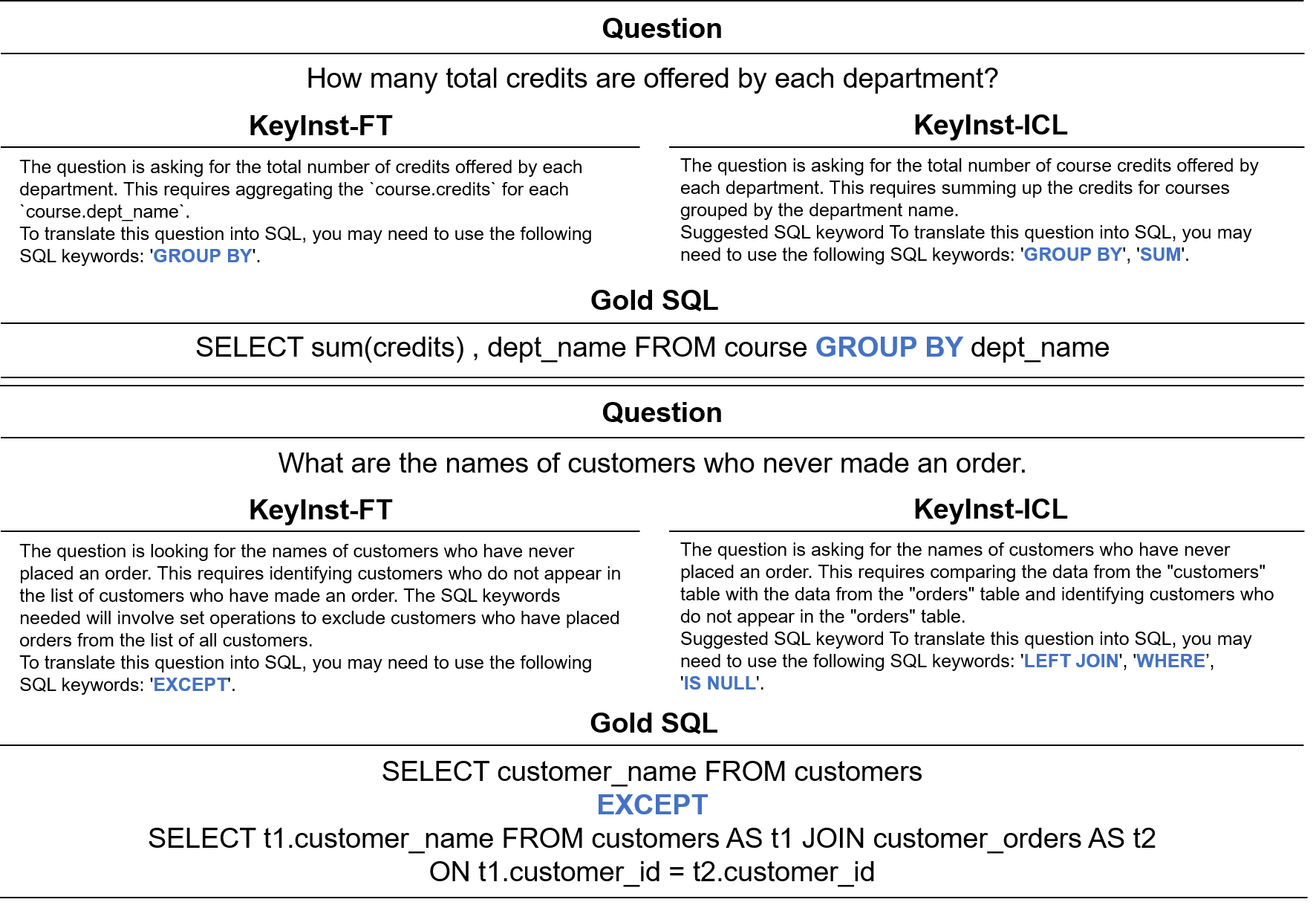}
	\caption{Examples of KeyInst-FT and KeyInst-ICL.}
    \label{fig: Examples of KeyInst-FT and KeyInst-IC}
\end{figure}

We propose two variants of KeyInst: KeyInst-FT and KeyInst-ICL KeyInst-FT is generated by a fine-tuned Llama3-8B model, while KeyInst-ICL is generated by guiding LLMs using In-Context learning. Our experiments demonstrate that KeyInst-FT performs better.
Examples of KeyInst-FT and KeyInst-ICL are provided in Figure \ref{fig: Examples of KeyInst-FT and KeyInst-IC}. These examples show that KeyInst-FT aligns more closely with the requirements of gold SQL. Specifically, KeyInst-FT consistently produces more accurate and contextually appropriate keyword suggestions.
This comparison highlights the advantage of fine-tuning models for specific tasks. 

\section{The usage of KeyInst}
\label{sec: The usage of KeyInst}
\begin{figure}[ht]
	\centering
	\includegraphics[width=0.8\linewidth,scale=1.00]{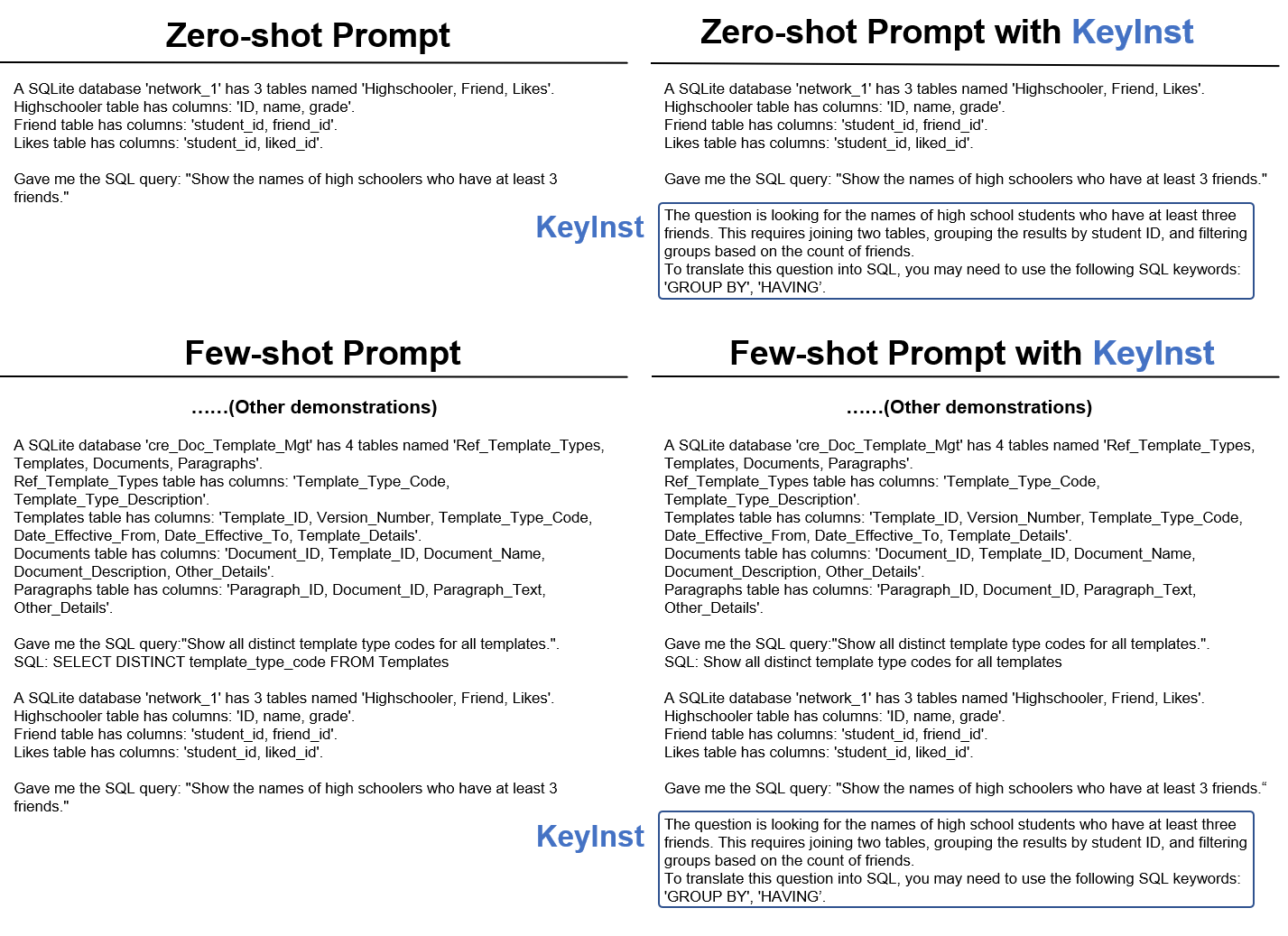}
	\caption{Examples of using KeyInst in zero-shot prompt and few-shot prompt.}
    \label{fig: The usage of KeyInst}
\end{figure}

KeyInst is represented as a single instruction, which gives it excellent compatibility and allows it to integrate with existing Text-to-SQL prompting methods seamlessly. Specifically, each KeyInst is tailored to the current Text-to-SQL task. To use it, we place it with the current Text-to-SQL task, typically at the end of the prompt. Figure \ref{fig: The usage of KeyInst} shows examples of using KeyInst. Note that in the few-shot prompt, we did not add KeyInst to each demonstrations, it is solely intended for the current Text-to-SQL task.

\section{Performance of LLMs with KeyInst on StrucQL}
\label{sec: Performance of LLMs with KeyInst on StrucQL}
We evaluated the performance of various LLMs on StrucQL after using KeyInst (KeyInst-FT). The results in Table \ref{tab: Performance of LLMs with KeyInst on StrucQL} show that KeyInst is a simple and effective method that significantly enhances the SQL formulation performance of LLMs. Notably, Llama3-70B with KeyInst is only 1.1\% behind GPT-4 with KeyInst.
\begin{table}[ht]
    \centering
    \resizebox{0.58\linewidth}{!}{
    \begin{tabular}{lccccc}
    \toprule[1pt]
    \textbf{Type} & \textbf{Gemma-7B} & \textbf{Llama3-8B} & \textbf{Llama3-70B} & \textbf{Claude3} & \textbf{GPT4}\\
    \midrule[0.5pt]   
    \multicolumn{6}{c}{\textit{Zero-shot}}\\
    \midrule[0.5pt]
    GROUP BY & 58.7    & 69.3    & \textbf{78. 0}  & 73.3   & 76.7   \\
    HAVING   & 52.7    & 74.0    & 82.0            & 85.3   & \textbf{86.0}   \\
    ORDER BY & 66.7    & 74.7    & 90.7            & 92.7    & \textbf{94.0}    \\
    LIMIT    & 51.3    & 76.0    & 80.7    & 77.3   & \textbf{88.7}   \\
    EXCEPT   & 41.3    & 63.3    & 64.7      & 70.7    &\textbf{ 72.7}    \\
    INTERSECT& 42.7    & 57.3    & 61.3   & \textbf{73.3}   & 72.7   \\
    UNION    & 30.7    & 38.7    & 47.3  & 55.3   & \textbf{57.3}  \\
    \midrule[0.5pt]
    Overall  & 49.1   & 63.9   & 72.0    & 75.4   & \textbf{78.3} \\

    \midrule[0.5pt]
    \midrule[0.5pt]
    
    \multicolumn{6}{c}{\textit{Zero-shot with KeyInst-FT}}\\
    \midrule[0.5pt]
    GROUP BY & 61.3    & 74.5    & \textbf{82.0}    & 78.0   & 78.7   \\
    HAVING   & 58.0    & 80.0    & 82.0    & \textbf{88.0}   & 86.0   \\
    ORDER BY & 71.3    & 88.0    & 92.7    & 93.3   & \textbf{94.0}    \\
    LIMIT    & 60.7    & 82.7    & 86.0    & 80.7   & \textbf{89.3 } \\
    EXCEPT   & 45.3    & 69.3    & 86.7    & \textbf{86.7}   & 83.3    \\
    INTERSECT& 56.7    & 81.3    & 84.7    & 85.3   & \textbf{85.3}   \\
    UNION    & 44.7    & 62.0    & 68.7    & \textbf{76.7}   & 73.3  \\
    \midrule[0.5pt]
    Overall  & 56.9    & 76.9    & 83.2    & 84.1   & \textbf{84.3} \\

    \bottomrule[1pt]
    \end{tabular}
    }
    \caption{Execution accuracy results for all compared LLMs on StrucQL after using KeyInst-FT.}
    \label{tab: Performance of LLMs with KeyInst on StrucQL}
\end{table}

\end{document}